\documentclass[a4paper]{article}
\usepackage{graphicx}
\usepackage{amsfonts}
\usepackage{tablefootnote}
\usepackage[dvipsnames]{xcolor}
\usepackage[utf8]{inputenc}
\usepackage[hang,flushmargin,bottom]{footmisc}
\usepackage{dblfloatfix}    
\usepackage{amsmath}
\usepackage{hyperref}
\usepackage{array}
\usepackage{amsthm}
\usepackage[skins]{tcolorbox}
\usepackage{tikztensor}
\usepackage{extramath}
\usepackage{multirow}
\usepackage{hhline}
\usepackage{adjustbox}
\usepackage{authblk}

\definecolor{kulblue}{RGB}{0,85,165}
\tikzset{fancy/.style={inner color=kulblue!5!white,%
outer color=kulblue!20!white,%
fill faces,%
opacity=0.95}}
\colorlet{kulblue20}{kulblue!20!white}
\colorlet{kulblue70}{kulblue!70!white}
\colorlet{kulblue30}{kulblue!30!black}
\colorlet{kulblue60}{kulblue!60!black}
\colorlet{kulblue90}{kulblue!90!black}

\title{Early soft and flexible fusion of EEG and fMRI \\ via tensor decompositions}

\vspace{5mm}

\author[1]{Christos Chatzichristos}
\author[2]{Eleftherios Kofidis}
\author[1,3]{Lieven De Lathauwer}
\author[4]{Sergios Theodoridis}
\author[1]{Sabine Van Huffel}

\affil[1] {KU Leuven, Department of Electrical Engineering (ESAT), STADIUS Center, Leuven, Belgium}

\affil[2]{Dept. of Statistics and Insurance Science, University of Piraeus, Greece}

\affil[3] {KU Leuven Kulak, Engineering and Technology, Kortrijk, Belgium}

\affil[4] {Chinese University of Hong Kong, Shenzhen, China}

\vspace{-4mm}

\begin{document}

\maketitle

%
%
\begin{abstract}
Data fusion refers to the joint analysis of multiple datasets which provide complementary views of the same task. In this preprint, the problem of jointly analyzing electroencephalography (EEG) and functional Magnetic Resonance Imaging (fMRI) data is considered. Jointly analyzing EEG and fMRI measurements is highly beneficial for studying brain function because these modalities have complementary spatiotemporal resolution: EEG offers good temporal resolution while fMRI is better in its spatial resolution. The fusion methods reported so far ignore the underlying multi-way nature of the data in at least one of the modalities and/or rely on very strong assumptions about the relation of the two datasets. In this preprint, these two points are addressed by adopting for the first time tensor models in the two modalities while also exploring double coupled tensor decompositions and by following soft and flexible coupling approaches to implement the multi-modal analysis. To cope with the Event Related Potential (ERP) variability in EEG, the PARAFAC2 model is adopted. The results obtained are compared against those of parallel Independent Component Analysis (ICA) and hard coupling alternatives in both simulated and real data. Our results confirm the superiority of tensorial methods over methods based on ICA. In scenarios that do not meet the assumptions underlying hard coupling, the advantage of soft and flexible coupled decompositions is clearly demonstrated.
 \end{abstract}

\maketitle

\vspace{-1mm}
\section{Introduction} 
\label{sec:intro}
In an attempt to better understand a system as complex as the human brain, multimodal measurements can be beneficial since they are able to provide information on complementary aspects of the same system. Through jointly analyzing the data resulting from different modalities, their individual advantages may be exploited and at the same time some of their disadvantages can be mitigated~\cite{2015_Lahat_Multimodal,2015_Adali_Fusiona}. In this way, a more accurate localization of the activated brain areas can be performed.

Two of the most commonly used modalities for monitoring the brain activity are the electroencephalography (EEG) and the functional Magnetic Resonance Imaging (fMRI). fMRI is a noninvasive brain imaging technique, which indirectly studies brain activity by measuring fluctuations of the blood-oxygen-level dependent (BOLD) signal~\cite{2008_lindquist_statistical}. The first BOLD fluctuation occurs roughly 2--3 seconds after the onset of the neural activity, when the oxygen-rich (oxygenated) blood starts displacing the oxygen-depleted (deoxygenated) blood. This rises to a peak after 4–6 seconds, before falling back to the baseline (and typically undershooting slightly).
The time course of the BOLD signal corresponding to a transient neural activity is called the Haemodynamic Response Function (HRF). Although fMRI has a high spatial resolution, often at the millimeter scale, it is a ``delayed'' measure of the brain activity, with its temporal resolution being limited by the repetition time of the scanner (TR), usually of the order of seconds~\cite{2008_lindquist_statistical}.

EEG provides information with respect to the neural electrical activity in the brain as a function of time. This is done via the use of multiple electrodes that are placed at certain locations over the scalp or (in more rare cases) over the cortex under the skull. The EEG signal results from the electrical measurement of the neuronal activation, realized through the movement of charged ions at the junction between the synapses of (the dendrites of) the neurons. This provides a more direct measure of the neuronal activity compared to fMRI (sensitive to millisecond changes in neural processing) and hence a better temporal resolution. However, EEG has poor spatial resolution, limited by the number of electrodes employed and the resistive properties of the extra-cerebral tissues. Furthermore, due to the fact that electrodes are more sensitive to neural activations that occur closer to the scalp, the determination of the exact location of activations that take place in deeper areas is more challenging~\cite{2007_sanei_eeg}. The complementary nature of their spatiotemporal resolutions motivates the fusion of EEG and fMRI towards a better localization of the brain activity, both in time and space~\cite{2015_Karahan_spacecoupling,2017_schward_thesis}.

Data fusion generally refers to the analysis of several datasets in a way that they interact and inform each other. Different types of fusion can be realized~\cite{2015_Lahat_Multimodal,2015_Karahan_spacecoupling,2017_ramachandram_typefusion} but generally the definition may differ with regard to the degree of generality and also
depending on the specific research areas~\cite{2015_cocchi_fusion}. Different types of applications, involving diverse sets of inter-related data, have been proposed. These include metabolomics~\cite{2007_Acar_metabolomics}, array processing~\cite{2013_sorensen_array}, sentiment analysis~\cite{2013_zadeh_sentiment},  multidimensional harmonic retrieval~\cite{2013_sorensen_harmonics1,2013_sorensen_harmonics2}, link prediction~\cite{2013_ermis_links} and, of course, biomedical applications~\cite{2015_Lahat_Multimodal,2015_Karahan_spacecoupling,2017_ramachandram_typefusion,2008_calhoun_ica,2015_Adali_Fusiona,2015_Adali_Fusionb} among many others. Fusion of EEG and fMRI data is expected to be of practical value given their complementary nature as described above.

\subsection{Categorization of data fusion}

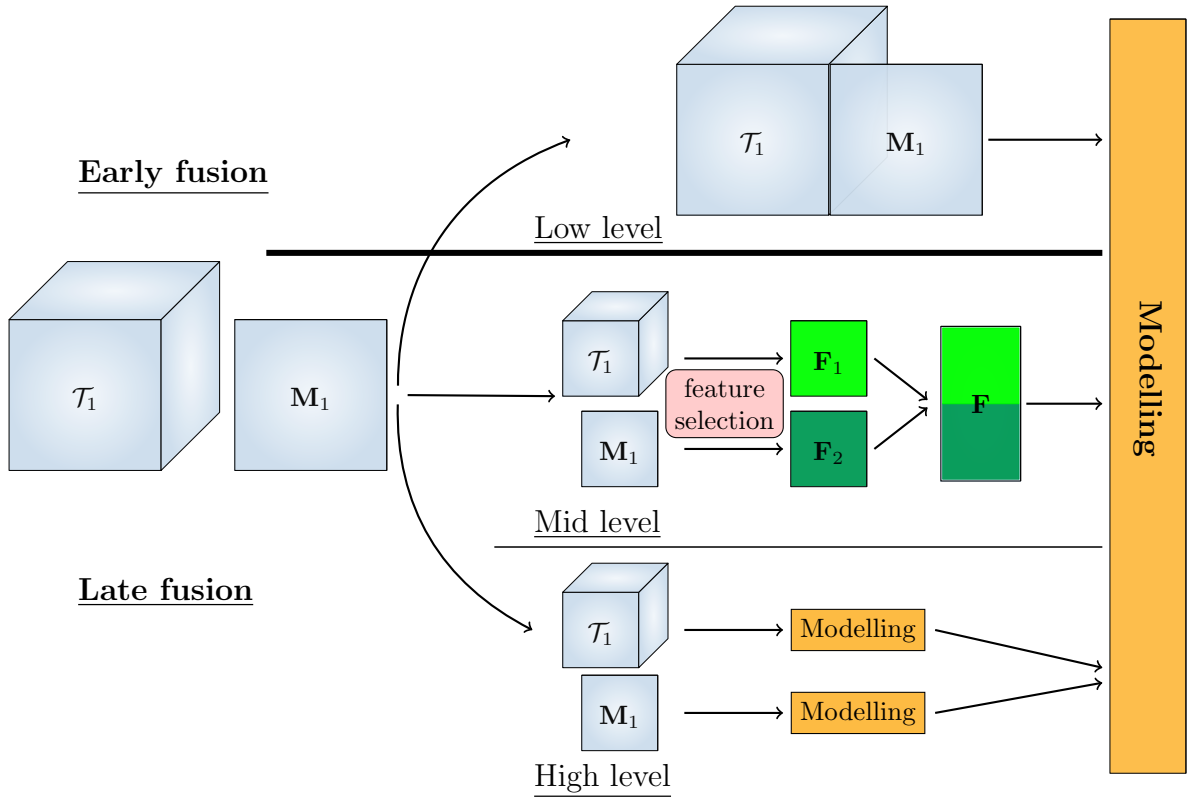
\begin{figure}
\hspace*{-1.5cm}
\begin{tikzpicture}[all tensors/.style={dim={2,2,2}, fancy}, node distance=0.4cm, chain]

\node at (-2,0) [tensor] (t) {$\ten{T}_1$};
\node [frontal matrix, dim={2,2},right=0.2cm of t, 2d] (m) {$\mat{M}_1$};
\node [right of = m, node distance = .75cm] (X1) {};
\node at (5,3) [tensor] (t2) {$\ten{T}_1$};
\node [frontal matrix, dim={2,2}, right=-0.75cm of t2,1d] (m2) {$\mat{M}_1$};
\draw [->,thick]  (X1) to [bend left=35] node[above] {} (4,3);
\draw [->,thick]  (9.5,3) -- (11,3) ;

\draw [line width=0.8mm]  (0,1.5) -- (11,1.5) ;
\draw [line width=0.2mm]  (3,-2.4) -- (11,-2.4) ;

\draw [->,thick]  (X1) to node {} (3.8,-0.4);
\node at (3.5,0.1) [tensor,dim={1,1,1}] (t3) {$\ten{T}_1$};
\node at (3.75,-1.1)[frontal matrix, dim={1,1}, 1d] (m3) {$\mat{M}_1$};
\node at (6.5,0.1)[frontal matrix, dim={1,1}, front fill=green, 2d] (m4) {$\mat{f}_1$};
\node at (6.5,-1.1)[frontal matrix, dim={1,1}, front fill=ForestGreen, 2d] (m5) {$\mat{f}_2$};
\node at (8.475,-0.5)[frontal matrix, dim={2.05,1.05}, 1d] (m6) {};
\node at (8.5,-1)[frontal matrix, dim={1,1}, front fill=ForestGreen, color=ForestGreen, 2d] (m6) {};
\node at (8.5,0)[frontal matrix, dim={1,1}, front
fill=green,color=green, 2d] (m7) {};
\node[text width=3cm] at (8.75, -0.5) {\textbf{F}};

\draw [->,thick]  (5.5,0.1) -- (6.8,0.1) ;
\node at (4.85,-0.5) [align=center,draw,rounded corners,fill=red!20] {feature\\selection};
\draw [->,thick]  (5.5,-1.1) -- (6.8,-1.1) ;
\draw [->,thick]  (8,0.1) -- (8.7,-0.45) ;
\draw [->,thick]  (8,-1.1) -- (8.7,-0.55) ;
\draw [->,thick]  (10,-0.5) -- (11,-0.5) ;

\draw [->,thick]  (X1) to [bend right=32] node[above] {} (3.5,-3.5);
\node at (3.5,-3.5) [tensor,dim={1,1,1}] (t3) {$\ten{T}_1$};
\node at (3.75,-4.6)[frontal matrix, dim={1,1}, 1d] (m3) {$\mat{M}_1$};
\node at (6.5,-3.5)[align=center, draw,fill=Dandelion] {Modelling};
\node at (6.5,-4.6)[align=center,draw,fill=Dandelion] {Modelling};
\draw [->,thick]  (5.5,-3.5) -- (6.8,-3.5) ;
\draw [->,thick]  (5.5,-4.6) -- (6.8,-4.6) ;
\draw [->,thick]  (8.8,-3.5) -- (11,-4) ;
\draw [->,thick]  (8.8,-4.6) -- (11,-4.2) ;

\node at (10.7,-0.4)[frontal matrix, dim={10,1}, front
fill=Dandelion, 2d] (m10) {};
\node[align=center,font=\large,rotate=270] at (11.2,1) {\textbf{Modelling}};

\node[align=center,font=\large] at (-3,2.5) {\textbf{\underline{Early fusion}}};
\node[align=center,font=\large] at (-3,-3) {\textbf{\underline{Late fusion}}};
\node[align=center,font=\large] at (3,1.8) {\underline{Low level}};
\node[align=center,font=\large] at (3,-2.1) {\underline{Mid level}};
\node[align=center,font=\large] at (3,-5.5) {\underline{High level}};
\end{tikzpicture}
\caption{Different types of data fusion approaches.}
\label{fusionlevel}
\end{figure}

Data fusion techniques can be categorized in various ways. The main categorizations are based on a) the level where the fusion is performed and b) the way the fusion is performed (Fig.~\ref{fusionlevel}). Two main levels and 2 sub-levels have been defined and become a reference classification~\cite{2017_ramachandram_typefusion,2019_hall_typefusion}, namely, ``early''/low-level fusion and ``late'' fusion, which is subdivided into mid-level fusion and high-level fusion. In ``early''/low-level fusion (or observational level), raw datasets (or blocks of data) are used. Mid-level (features level or state-vector level) fusion is considered when the data fusion methods operate on features extracted from each dataset separately, so, instead of using raw data for modelling the task at hand (e.g., classifying), features of the data are used. The high-level (decision/information level) fusion methods model each dataset separately and only decisions (model outcome) from processing of each data block are fused.

The categorization based on the way the fusion is performed is two-way. The earliest approaches for fusion of fMRI and EEG (and a large number of recent ones, e.g., ~\cite{2015_ferdowsi_new}) are essentially ``integrative'' in nature. The rationale behind these methods is to employ objective functions for decomposition of the fMRI signal with constraints based on information from EEG (or vice versa). Recently, the emphasis has been turned to ``true'' fusion, e.g.,~\cite{2016_hunyadi_fusion,2017_acar_acmtf1,2017_acar_acmtf2,2017_Eyndhoven_HRF,2004_Martinez_PLS,2015_Karahan_spacecoupling}, where the decomposition of the data from each modality can influence the other using all the common information that may exist. During optimization, the factors, which have been identified as shared, are appropriately ``coupled'' and thus a bridge between the two modalities is established. For a detailed literature review of such methods, the interested reader is reffered to~\cite{2018_kofidis_partially,2015_Adali_Fusiona}.

\subsection{Fusion of EEG and fMRI}

Multivariate bilinear (i.e., matrix-based) methods, mainly based on Independent Component Analysis (ICA)~\cite{2006_calhoun_jica_pica,2012_mijovic_whyhowjica,2014_swinnen_jica,2015_hunyadi_parallel} and relying on the concatenation of different modes, have been, up to recently,  the state of the art for jointly analyzing EEG and fMRI. However, by definition such methods fall short in exploiting the inherently multi-way nature of these data. fMRI and EEG datasets are inherently multi-dimensional, comprising information in time and along different voxels or channels, subjects, trials, etc. For EEG, in order to better exploit the information, the signal can be expanded in additional dimensions, e.g. through incorporating spectral features by computing a wavelet transform of the EEG data or using the segment/Event Related Potential (ERP) mode (ERP is the response to a specific sensory, cognitive, or motor stimulus)~\cite{2015_cong_tensorseeg}. This multi-dimensional nature of the EEG and fMRI datasets points to the adoption of tensor (multi-linear) models instead of the bi-linear ones. Several tensor decomposition methods have been applied in fMRI and EEG Blind Source Separation (BSS), including Canonical Polyadic Decomposition (CPD) or Parallel Factor Analysis (PARAFAC)~\cite{2004_andersen_structure-seeking,2007_Acar_cpdeeg}, and its generalizations known as PARAFAC2~\cite{2017_chatzichristos_BTD2,2017_Loukianos_PFAC2} and Block Term Decomposition (BTD)~\cite{2012_de_lathauwer_block,2019_chatzichristos_journal}. 

The representations that are possible with tensor models can a) improve the ability of extracting spatiotemporal modes of interest~\cite{2004_andersen_structure-seeking,2007_stegeman_comparing,2013_helwig_critique}, b) facilitate neurophysiologically meaningful interpretations~\cite{2004_andersen_structure-seeking}, and c) produce unique (modulo scaling and permutation ambiguities) representations under mild conditions~\cite{2000_sidiropoulos_uniqueness}. Those mild conditions can be even more relaxed in the case of coupled tensor decompositions than their single-tensor counterparts. It has been demonstrated that coupling through one or more common factors that are shared among tensors can ensure uniqueness beyond what is possible
when considering separate decompositions~\cite{2015_sorensen_coupled}. Moreover, tensorial methods are able to make predictions more robustly in the presence of noise, compared to their two-way counterparts~\cite{2004_andersen_structure-seeking,2017_sidiropoulos_reviewtensor,2019_chatzichristos_journal}. It should be noted that the biomedical data are usually highly corrupted by noise~\cite{2004_andersen_structure-seeking}. 

Various ways to realize the coupling have been proposed, depending on the coupled mode: a) coupling in the spatial domain with the use of the so-called lead-field matrix, which summarizes the volume conduction effects in the head (by transforming the 2D spatial information of the EEG to the 3D spatial information of the fMRI) ~\cite{2015_Karahan_spacecoupling}, b) coupling in the time domain using the convolution of the EEG time course with an HRF~\cite{2004_Martinez_PLS}, and c) coupling in the subject domain, using the assumption that the same neural processes are reflected in both modalities with the same covariation~\cite{2006_calhoun_jica,2014_swinnen_jica,2016_hunyadi_fusion}.

Heterogeneity in the datasets is also manifested in the models used to represent them. In the EEG-fMRI fusion example, classical approaches adopt a space (channels)  $\times$ time  $\times$ frequency/ERP tensor model for EEG (for the single-subject case) whereas the fMRI signal is commonly represented as a matrix with its dimensions corresponding to space (voxels)  $\times$  time. Their fusion relies on the coupling of the EEG tensor and the fMRI matrix along their common mode (in one of the ways described before). Thus, although the multi-way nature of EEG has been exploited in earlier fusion methods~\cite{2017_acar_acmtf1,2016_hunyadi_fusion}, it has been so far neglected for fMRI. Furthermore, those methods rely on preprocessing of the fMRI data using the General Linear Model (GLM) framework. A spatial map of interest (areas of activation) per subject is extracted from the fMRI data and all the spatial maps are stacked in a matrix (space $\times$ subjects), hence discarding the extra dimension of time and relying on Coupled Matrix Tensor Factorization (CMTF) to solve the joint BSS problem.\footnote{Advanced CMTF (ACMTF)~\cite{2017_acar_acmtf1,2017_acar_acmtf2} allows the presence of both shared and unshared components in the coupled factor(s) and provides a way to automatically determine them. Recently the uniqueness properties of such partially coupled decompositions have been studied~\cite{2018_kofidis_partially,2019_sorensen_coupledpartially}.} In the GLM framework, a canonical HRF is assumed to be known (and be invariant in space and among subjects), the expected signal changes are defined as regressors of interest in a multiple linear regression analysis and the estimated coefficients are tested against a null hypothesis. In the EEG and fMRI studies using GLM, the EEG signal (or part of it) is used as the regressor of interest. Intra- and inter-subject variability of HRF is known to exist~\cite{2004_handweker_bold}, hence a possible misspecification of the HRF may lead to biased estimates of widespread activity in the brain~\cite{2008_lindquist_statistical,2004_handweker_bold}. Moreover, the mismatch of the temporal characteristics of EEG and fMRI further limits the potential of GLM analysis~\cite{2015_hunyadi_parallel}. The use of the spatial maps of GLM categorizes such CMTF-based methods as late fusion~\cite{2017_ramachandram_typefusion}.

In all of the approaches that were previously described, the coupling between the corresponding modes is ``hard'', meaning that the shared factors are constrained to be equal in the two datasets (after any transformation applied, e.g., convolution with an HRF). Such an assumption is very restrictive, since it implies that the used transformation is valid for every area of the brain and any subject. In order to alleviate any problems caused in the modelling by the fact that the shared factors are forced to be the same between modalities, a ``softer'' assumption of similarity (or with similar properties), not necessarily of the strong equality, can be made instead~\cite{2014_seichepine_soft,2015_Farias_softMultimodal}. Furthermore, different methods can be used to account for a possible misspecification of the HRF. Constraining the HRF to a class of ``plausible'' waveforms and estimating the optimal one from the data itself has been proposed in~\cite{2017_Eyndhoven_HRF} for the single-subject case. Such approaches will be called ``flexible''. 

In this work, we investigate early~\cite{2017_ramachandram_typefusion} fusion of fMRI and EEG via soft (assuming similarity and not strong/hard equality) and flexible coupling. As explained previously, soft and flexible coupling are different ways to accommodate for a possible missmodelling of the HRF. Their main difference is that with soft coupling all the HRFs of the different subjects are assumed to be similar (and not equal) with an a-priori known HRF; while in the flexible approach only the model of the HRF is a-priori known and the variables of the model, which determine  the exact shape of the HRF, are estimated via optimization. In our approach, we want to  demonstrate the gains from: 

\begin{itemize}
    \item Using raw data instead of features (early fusion), omitting the GLM preprocessing step in an effort to fully exploit the information underlying the raw data~\cite{2017_ramachandram_typefusion,2015_Lahat_Multimodal}
    \item Exploiting the multi-way nature of both modalities either by multi-way tensors (when possible)  for both modalities or double CMTFs
    \item Using flexible and soft coupling models in order to alleviate the problem of mismodelling of the HRF. 
\end{itemize}

We also want to compare the flexible and soft coupling methods via simulated data. Furthermore, we propose an alternative modelling for the HRF, and we demonstrate the advantage of the proposed methods over methods based on ICA, hard coupling and uncoupled CPD per modality. 

\subsection{Notation}

Vectors, matrices and higher-order tensors are denoted by bold lower-case, upper-case and calligraphic upper-case letters, respectively. For a matrix $\boldsymbol{A}$, $\boldsymbol{A}^\top$ and $\boldsymbol{A}^{\dag}$ denote its transpose and pseudo-inverse, respectively. An entry of a vector $\boldsymbol{a}$, a matrix $\boldsymbol{A}$, or a (3rd-order) tensor $\boldsymbol{\mathcal{A}}$ is denoted by $a_i$, $a_{i,j}$,  or $a_{i,j,k}$, respectively. Matlab notation is used to denote a column of a matrix $\mathbf{A}$, namely $\mathbf{A}(:,j)$ is its $j$th column. $\boldsymbol{I}_m$ is the $m$th-order identity matrix and $\boldsymbol{1}_{m}$ denotes the $m\times 1$ vector of all ones. The symbols $\otimes$ and $*$ denote the Kronecker and the Hadamard (elementwise) products, respectively. The column-wise Khatri--Rao product of two matrices, $\boldsymbol{A} \in \mathbb{R} ^{I\times R}$ and $\boldsymbol{B} \in \mathbb{R} ^{J\times R}$, is denoted by $\boldsymbol{A}\odot\boldsymbol{B}=\begin{bmatrix}\boldsymbol{a}_1\otimes \boldsymbol{b}_1, \boldsymbol{a}_2\otimes \boldsymbol{b}_2,\ldots, \boldsymbol{a}_R\otimes \boldsymbol{b}_R \end{bmatrix}$, with $\boldsymbol{a}_j,\boldsymbol{b}_j$ being the $j$th columns of $\boldsymbol{A},\boldsymbol{B}$, respectively. The outer product of two tensors is denoted by $\circ$. For an $N$th-order tensor, $\boldsymbol{\mathcal{A}} \in \mathbb{R} ^{I_1 \times I_2 \times \cdots \times I_N}$, $\boldsymbol{A}_{(n)}\in \mathbb{R} ^{I_n \times I_1I_2 \cdots I_{n-1}I_{n+1} \cdots I_N}$ is its mode--$n$ unfolded (matricized) version (whose rank is known as mode--$n$ rank), which results from mapping the tensor element with indices $(i_1,i_2,\ldots,i_N)$ to a matrix element $(i_n,j)$, with $j=1 + \sum_{k=1,k\neq n}^N [ ( i_k -1 ) J_k]$, $J_k=~\begin{cases} 
  1, \qquad \mathrm{for} \quad k=1 \: \mathrm{or} \: k=2 \: \mathrm{and} \:  n=1, \\ 
  \prod_{m=1,m\neq n}^{k-1}I_m, \qquad \mathrm{otherwise}.
  \end{cases}$ \\

\section{Methods}
\label{sec:theory}

\subsection{Canonical Polyadic Decomposition (CPD)}
CPD (or PARAFAC) \cite{2017_sidiropoulos_reviewtensor} approximates a 3rd-order tensor, $\boldsymbol{\mathcal{T}} \in \mathbb{R} ^{I_1 \times I_2\times I_3}$ (naturally extended to tensors of higher order), by a sum of $R$ rank-1 tensors,

\begin{equation}
\label{cpd1}
\boldsymbol{\mathcal{T}} \approx \sum_{r=1}^{R} \boldsymbol{a}_r \circ \boldsymbol{b}_r \circ \boldsymbol{c}_r
\end{equation}
Equivalently, for the $k$th frontal slice of $\boldsymbol{\mathcal{T}}$,
\begin{equation}
\label{cpd3}
\boldsymbol{T}_{k} \approx \boldsymbol{A} \boldsymbol{D}_k \boldsymbol{B}^\top, \quad k=1,2,\ldots,I_3
\end{equation}

\noindent where $\boldsymbol{A}=\begin{bmatrix}\boldsymbol{a}_1,\boldsymbol{a}_2,\ldots,\boldsymbol{a}_R\end{bmatrix}$, $\boldsymbol{B}$ and $\boldsymbol{C}$ are similarly defined matrices, and $\boldsymbol{D}_k$ is the diagonal matrix having the elements of the $k$th row of $\boldsymbol{C}$ on its diagonal. The main advantage of the CPD, besides its simplicity, is the fact that it is unique (up to permutation and scaling) under mild conditions~\cite{2017_sidiropoulos_reviewtensor}. Uniqueness of CPD is crucial to its application in BSS problems. Its performance is, however, largely dependent on the correct estimation of the tensor rank, $R$. Several heuristic methods have been proposed for the latter problem~\cite{2003_bro_new}.

\subsection{PARAFAC2}
PARAFAC2~\cite{2017_sidiropoulos_reviewtensor} differs from CPD in that strict multilinearity is no longer a requirement. CPD applies the same factors across all the different slices, whereas PARAFAC2 relaxes this constraint and allows variation across one of the modes (in terms of the values and/or the size of the corresponding factor matrix). For this reason, PARAFAC2 is not a tensor model in the strict sense as it can represent both regular tensors, with weaker constraints than CPD, as well as irregular tensors (collections of matrices of different dimensions) with size variations along one of the modes. It can be written in terms of the (here frontal) slices of the tensor $\boldsymbol{\mathcal{T}}$ as

\begin{equation}
\label{parafac22}
\boldsymbol{T}_{k} \approx \boldsymbol{A}_k \boldsymbol{D}_k \boldsymbol{B}^\top, \quad k=1,2,\ldots,I_3 ,
\end{equation}

\noindent with $\boldsymbol{A}_k$ being different for different $k$'s. This type of decomposition is clearly non-unique. Thus, in order to allow for uniqueness, it has been proposed to add the constraint that the cross products $\boldsymbol{A}_k^\top \boldsymbol{A}_k$ be constant over $k$. This has been shown~\cite{1999_kiers} to be equivalent to setting $\boldsymbol{A}_k=\boldsymbol{P}_k\boldsymbol{F}$, where the $R \times R$ matrix $\boldsymbol{F}$ is the same for all slices, while the variability is represented by the columnwise orthonormal $I_2 \times R$ matrix $\boldsymbol{P}_k$. Under this constraint, one has to fit the equivalent model

\vspace{-1mm}
\begin{equation}
\label{parafac23}
\boldsymbol{P}_k^\top\boldsymbol{T}_{k} \approx \boldsymbol{F} \boldsymbol{D}_k \boldsymbol{B}^\top, \quad k=1,2,\ldots,I_3.
\end{equation}
\vspace{-2mm}

As shown in~\cite{1999_kiers}, $\boldsymbol{P}_k$ can be computed as  $\boldsymbol{P}_k=\boldsymbol{V}_k\boldsymbol{U}_k^\top$, where $\boldsymbol{U}_k$ and $\boldsymbol{V}_k$ are the left and right singular matrices of $\boldsymbol{F} \boldsymbol{D}_k \boldsymbol{B}^\top\boldsymbol{T}_{k}^\top$. As can be seen from Eq.~(\ref{parafac23}), the problem of fitting PARAFAC2 has been transformed into that of fitting a CPD model with transformed data. Applications of PARAFAC2 in fMRI and EEG analysis include~\cite{2015_ferdowsi_new,2017_chatzichristos_BTD2} and ~\cite{2017_Loukianos_PFAC2}, respectively. 

\subsection{ICA-based methods}
  Classical approaches for jointly analyzing fMRI and EEG include Joint Independent Component Analysis (JICA) (using one~\cite{2006_calhoun_jica,2006_calhoun_jica_pica} or multiple~\cite{2014_swinnen_jica} electrodes for EEG), and Parallel ICA~\cite{2015_hunyadi_parallel,2006_calhoun_jica_pica}. JICA jointly analyzes data from the same subjects from both modalities simultaneously. To achieve this, it uses the features derived from the first-level analysis of fMRI (spatial maps) and the averaged ERP epochs of EEG, hence JICA is also classified as a late fusion model. JICA assumes that a stronger ERP yields a stronger BOLD fluctuation in the same area (and vice versa), which supports the common assumption of having the same linear mixing system in the two modalities (in the subjects domain). Furthermore, each pair of coupled components is assumed to be dependent between the modalities and at the same time statistically independent of the rest of the components~\cite{2012_mijovic_whyhowjica}. 
 Parallel ICA first identifies components separately for each modality, performing a temporal ICA in EEG and a spatial ICA in fMRI. In a second step, the corresponding extracted components are identified based on their correlation in the temporal domain. Parallel ICA can be performed either at a single-subject level~\cite{2015_hunyadi_parallel} or at a multi-subject level using Group ICA~\cite{2010_lei_steff}.
 
\subsection{Modelling of the HRF}

As mentioned in the introduction, the GLM framework is most commonly adopted in fMRI analysis. Analysis within the GLM is rooted in the simple assumption that the variance in the fMRI BOLD signal can be modeled by the convolution of a (assumed to be known) HRF with the event/stimulus. The haemodynamic response is composite and nonlinear, resulting from the neuronal and vascular changes, which is known to vary among different subjects as well as among different areas of the same brain (inter-subject and intra-subject variability)~\cite{2004_handweker_bold}.

GLM-based methods explicitly need an estimate of the functional shape of the HRF to infer the expected activation pattern from the experimental task. Among the different available models for the HRF, the one that is more widely used is the model based on the two Gamma distributions~\cite{2008_lindquist_statistical,2004_handweker_bold}, usually referred to as double Gamma HRF model:
\begin{equation}
 H(t,z)=\Gamma^{-1}(z_{(1)}) z_{(2)}^{z_{(1)}} t^{z_{(1)}-1} \mathrm{e} ^{-z_{(2)}t}  - z_{(3)} \Gamma^{-1}(z_{(4)}) z_{(5)}^{z_{(4)}} t^{z_{(4)}-1} \mathrm{e} ^{-z_{(5)}t},
\end{equation}

\noindent where $\Gamma(\cdot)$ is the Gamma function, $\Gamma(\cdot)^{-n}=1/\Gamma(\cdot)^{n}$, and $z_{(1,2,3,4,5)}$ are the parameters that control the functional shape of the HRF. The values $z_{(1)}=~6$, $z_{(2)}~=1, z_{(3)}=~\frac{1}{6},z_{(4)}=16,z_{(5)}=1$ are used to generate the canonical HRF used in GLM.

Several other models have been proposed, such as the methods based on the cosine function~\cite{2004_zarahn_hrf}, radial bases~\cite{2004_riera_hrf}, and spectral basis function~\cite{2002_liao_hrf}. Furthermore, neuro-physiologically informed non-linear models of the HRF have been proposed, describing the dynamic changes in deoxyhemoglobin content as a function of blood oxygenation and blood volume~\cite{1998_buxton_dynamics,2004_buxton_modeling}, with a model of the blood flow dynamics during brain activation, where neuronal activity is approximated by the stimulus/task input scaled by a factor called neural efficiency, in the so-called ``balloon'' model. However, it must be pointed out that the models exhibit differences both in capturing the evoked changes of the HRF as well as in the number of parameters used to model the HRF~\cite{2009_lindquist_hrf}.

In this work, a new lighter model for the functional shape of the HRF will be tested, based on the Lennard-Jones potential~\cite{2018_LennardJones_wiki}. The latter is used in physics to model the repulsive and attractive forces between neutral atoms or
molecules. Due to its computational simplicity, the Lennard-Jones potential is used extensively in computer simulations even though more accurate potentials exist. This light model will be used in view of the smaller number of parameters used and the smoother partial derivatives which will be used during the optimization\footnote{A detailed description and motivation of the use of the Lennard-Jones potential along with a fit analysis with real data can be found in~\cite{2020_morante_hrf}. }. The Lennard-Jones model (as it will be henceforth referred to) is defined over the non-negative real numbers and can be expressed as:
\begin{equation}
 H(t,z)=\Gamma^{-3}(z_{(1)} t) - z_{(2)} \Gamma^{-6}(z_{(3)} t)
\end{equation}

\noindent where $z_{(1,2,3)}$ are the parameters that control the functional shape of the HRF. Therefore, it can be noted that the proposed model only has three such parameters, compared to the five parameters of the double Gamma distribution model above. 

Its time derivative can be obtained as follows:
\begin{equation}
   \frac{\partial H}{\partial t}=-3 z_{(1)} \Gamma^{-3}(z_{(1)}t)\psi_{0}(z_{(1)}t)+6z_{(2)} z_{(3)}\Gamma^{-6}(z_{(3)}t)\psi_{0}(z_{(3)}t),
\end{equation}
	
\noindent where $\psi_{0}$ is the polygamma function~\cite{2020_LennardJones_polygamma} of order zero, also called digamma function.
Furthermore, the partial derivatives of the function with respect to each of the parameters are given as:

\textbf{Parameter $z_{(1)}$:}
\begin{equation}
	\frac{\partial H}{\partial z_{(1)}}=-3 t\Gamma^{-3}(z_{(1)}t)\psi_{0}(z_{(1)}t)
\end{equation}

\textbf{Parameter $z_{(2)}$:}
\begin{equation}
	\frac{\partial H}{\partial z_{(2)}}=-\Gamma^{-6}(z_{(3)} t)
	,
\end{equation}

\textbf{Parameter $z_{(3)}$:}
\begin{equation}
	\frac{\partial H}{\partial z_{(3)}}=6 z_{(2)} t  \Gamma^{-6}(z_{(3)}t)\psi_{0}(z_{(3)}t)
\end{equation}

Note that the partial derivatives that will be used in the non-linear least squares (nls) optimization framework, are much simpler than the corresponding derivatives of the double Gamma HRF model. 

\section{Soft-Coupled Tensor Decompositions}

Coupling through equality (hard coupling), which is used both in CMTF-based methods~\cite{2016_hunyadi_fusion,2017_acar_acmtf1,2017_acar_acmtf2} and in JICA~\cite{2012_mijovic_whyhowjica,2014_swinnen_jica,2006_calhoun_jica} approaches, arises from the assumption that the neural sources are reflected, exactly with the same power, in both modalities; however this is restrictive. Even if the exact equality and the independence assumptions, used by JICA, are valid, still the result of the first-level analysis of fMRI (used as an initial step~\cite{2016_hunyadi_fusion,2017_acar_acmtf1,2017_acar_acmtf2,2006_calhoun_jica,2012_mijovic_whyhowjica,2014_swinnen_jica}) is not taking into account the complementary information of EEG. Furthermore, as reported in~\cite{2012_mijovic_whyhowjica}, the result obtained with JICA is mostly influenced by the quality of the ERPs (EEG) and less by the fMRI data. This may indicate that the preprocessing of the fMRI with GLM may fail to retrieve all the information ``hidden'' in the raw fMRI data, due to the constraints of GLM~\cite{2008_lindquist_statistical}. 

\begin{figure}
    \centering
    \includegraphics[width=.6\linewidth]{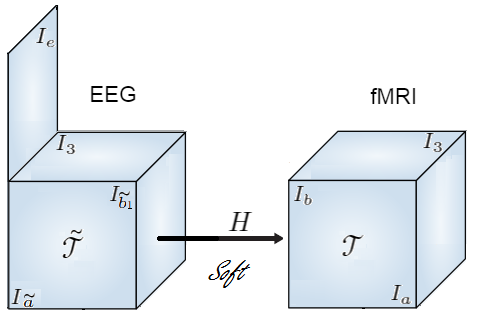}
    \caption{Schematic representation of coupled CPDs with ``soft'' coupling.}
    \label{softcoupl}
\end{figure}

We propose a framework for early fusion of fMRI and EEG using coupled CPD with soft coupling~\cite{2014_seichepine_soft}, which means similarity and not exact equality (Fig.~\ref{softcoupl}). Fusion based on raw data, though potentially more challenging, may allow better inference~\cite{2017_ramachandram_typefusion}. The coupling could be attempted in any of the modes, depending on the problem at hand.

Considering the 3rd-order fMRI tensor, $\boldsymbol{\mathcal{T}} \in \mathbb{R} ^{I_a \times I_b \times I_3}$ (space $\times$ time $\times$ subjects), and the 4th-order EEG tensor, $\boldsymbol{\mathcal{\tilde{T}}} \in \mathbb{R} ^{I_e \times I_{\tilde{a}}\times I_{\tilde{b}} \times I_3}$ (ERPs/frequency $\times$ space $\times$ trials amplitude $\times$ subjects). Their CPDs can be written as $\boldsymbol{T}_k \approx \boldsymbol{A} \boldsymbol{D}_k \boldsymbol{B}^\top$ and $\boldsymbol{\tilde{T}}_{k(1)} \approx \boldsymbol{E}\boldsymbol{\tilde{D}}_k (\boldsymbol{\tilde{B}} \odot \boldsymbol{\tilde{A}} )^\top $, respectively, with $\boldsymbol{\tilde{T}}_{k(1)}$ being the mode-1 matricization of $\boldsymbol{\mathcal{\tilde{T}}}_k=\boldsymbol{\mathcal{\tilde{T}}}(:,:,:,k)$~\cite{2017_chatzichristos_BTD2}. $\boldsymbol{A}=\begin{bmatrix}\boldsymbol{a}_1,\boldsymbol{a}_2,\ldots,\boldsymbol{a}_R\end{bmatrix}$ is a matrix that contains the weights of the $R$ spatial components ($I_a$ voxels), $\boldsymbol{B},\boldsymbol{C}$ contain the associated time courses $(I_b)$ and subject activation levels of fMRI $(I_3)$, respectively, and $\boldsymbol{D}_k$ is the diagonal matrix formed from the $k$th row of $\boldsymbol{C}$. For the EEG case, matrices $\boldsymbol{E},\boldsymbol{\tilde{A}},\boldsymbol{\tilde{B}},\boldsymbol{\tilde{C}}$ contain the weights of the  associated ERPs $(I_e)$, electrodes $(I_{\tilde{a}})$, trials amplitude $(I_{\tilde{b}})$ and the subject activation levels of EEG $(I_3)$, respectively, and $\boldsymbol{\tilde{D}}_k$ is the diagonal matrix formed from the $k$th row of $\boldsymbol{\tilde{C}}$. The proposed cost function to be minimized is given by

\begin{align}
\label{hatzi}
&\sum_{k=1}^{I_3} \| \boldsymbol{T}_k - \boldsymbol{A} \boldsymbol{D}_k \boldsymbol{B}^\top \|_F^2 
+ \sum_{k=1}^{I_3} \| \boldsymbol{\tilde{T}}_{k(1)} - \boldsymbol{E} \boldsymbol{\tilde{D}}_k (\boldsymbol{\tilde{B}} \odot \boldsymbol{\tilde{A}} )^\top\|_F^2  \nonumber \\ & + \lambda_A \| \boldsymbol{LA}_{1:R_c} - \boldsymbol{ \tilde{A}}_{1:R_c} \|_F^2
+ \lambda_B \| \boldsymbol{B}_{1:R_c} - \boldsymbol{H\tilde{B}}_{1:R_c} \|_F^2  \\ & + \lambda_C \| \boldsymbol{C}_{1:R_c} - \boldsymbol{\tilde{C}}_{1:R_c} \|_F^2 \nonumber,
\end{align}

\noindent with $\boldsymbol{L}$ being the lead-field matrix used for the EEG forward problem and $\boldsymbol{H}$ the matrix representing the convolution with the HRF and the down-sampling (due to the different sampling rate of the two modalities). The values of $\lambda$'s quantify the degree of coupling. It shall be noted that the weights of the different modalities are set to the unit, due to the fact that they have been both normalised to unity norm prior to the analysis (which is a really important preprocessing step). $R_c$ is the number of common components in the coupled mode(s), so there are $R-R_c$ and $\tilde{R}-R_c$ distinct components of fMRI and EEG, respectively. In this way, different model orders can be assigned to the decompositions of the modalities as long as the number of common components remains the same (without loss of generality, in (11), we assume that the common components are the first $R_c$ ones). 

As can be noted in Eq.~(11), the quadrilinear model of CPD selected for decomposing the EEG tensor assumes that every subject has exactly the same ERP, an assumption which is restrictive~\cite{2009_sur_erp} and can be relaxed with the adoption of PARAFAC2~\cite{2010_weis_parafac2,2009_sur_erp}, where $\boldsymbol{E}$ may vary with $k$. Thus, the CPD used for EEG can be replaced by PARAFAC2, with $\boldsymbol{E}_k=\boldsymbol{P}_k \boldsymbol{F}$ and $\boldsymbol{P}_k$ and $\boldsymbol{F}$ computed as in Section 2.2, and the cost function (11) becomes

\begin{align}
\label{hatzi2}
&\sum_{k=1}^{I_3} \| \boldsymbol{T}_k - \boldsymbol{A} \boldsymbol{D}_k \boldsymbol{B}^\top \|_F^2 
+ \sum_{k=1}^{I_3} \| \boldsymbol{P}_k^\top\boldsymbol{\tilde{T}}_{k(1)} - \boldsymbol{F} \boldsymbol{\tilde{D}}_k (\boldsymbol{\tilde{B}} \odot \boldsymbol{\tilde{A}} )^\top\|_F^2  \nonumber \\ & + \lambda_A \| \boldsymbol{LA}_{1:R_c} - \boldsymbol{ \tilde{A}}_{1:R_c} \|_F^2
+ \lambda_B \| \boldsymbol{B}_{1:R_c} - \boldsymbol{H\tilde{B}}_{1:R_c} \|_F^2  \\ & + \lambda_C \| \boldsymbol{C}_{1:R_c} - \boldsymbol{\tilde{C}}_{1:R_c} \|_F^2 \nonumber.
\end{align}

\section{Double Coupled Matrix \\ Tensor Factorization 
(DCMTF)}

\begin{figure}[!b]
    \centering
    \includegraphics[width=\linewidth]{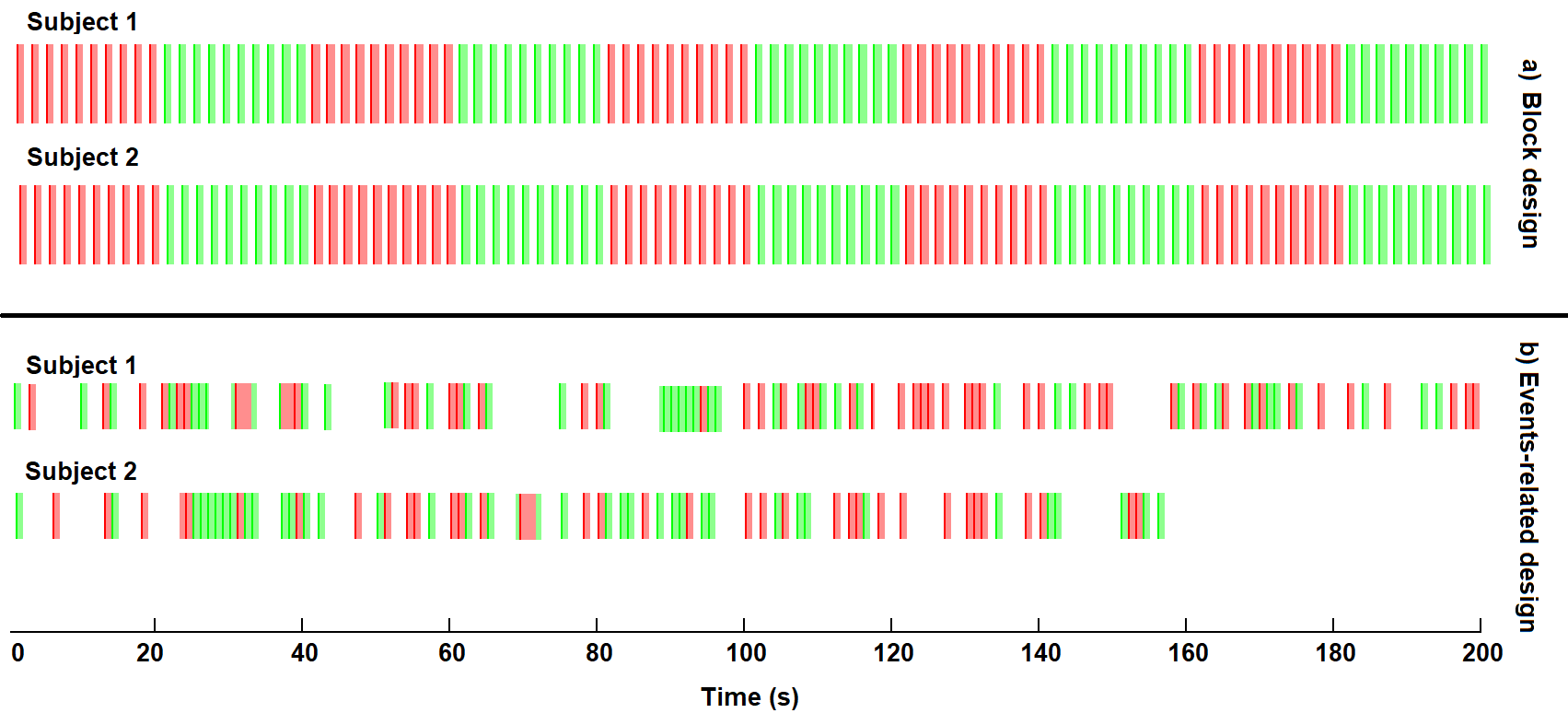}
    \caption{Types of experimental fMRI design: a) Block event design b) Event-related design.}
    \label{designs}
\end{figure}

As noted previously, the CPD model assumes multi-linearity for all the modes. The multi-linearity in the ERP/frequency mode can be relaxed with the use of PARAFAC2. Depending on the design of the experiment another assumption used, in order to stack different subject in a tensor, may be inaccurate. In task-related fMRI, currently two major classes of fMRI experimental designs exist: block designs and event-related designs~\cite{2008_lindquist_statistical,2009_Tie_designs}. In a blocked design, a condition is presented continuously for an extended time interval (block) to maintain cognitive engagement, with different task conditions usually alternating in time. The time course of the stimuli (both the sequence of the stimuli and the time intervals) remain stable among subjects (Fig.~\ref{designs}.a). In an event-related design, discrete and short-duration events are presented with randomized timing and order (both during the acquisition of a single subject but also among different subjects). Both designs have advantages and disadvantages. For example, block event design is more robust since relatively large BOLD signal changes with increased statistical power are detected. Moreover, it is statistically powerful and straightforward to analyze, in the sense that the exact shape of the HRF does not influence much the result of the analysis and hence can be assumed to be simple (equal to canonical) with smaller impact. On the other hand, the predictability of block design makes it inappropriate for some cognitive tasks, such as an `oddball' paradigm where a reaction to an unexpected stimulus is examined. Furthermore, it also increases the chance of low-frequency artifacts. Event-related design can detect transient variations in haemodynamic response and allows for the analysis of individual responses to trials. Furthermore, a study connected to the detection of a specific disease, e.g., seizure detection, follows a design similar to an event-related design, since a possible seizure onset can not be aligned among all subjects.

\begin{figure}
    \centering
    \includegraphics[width=.6\linewidth]{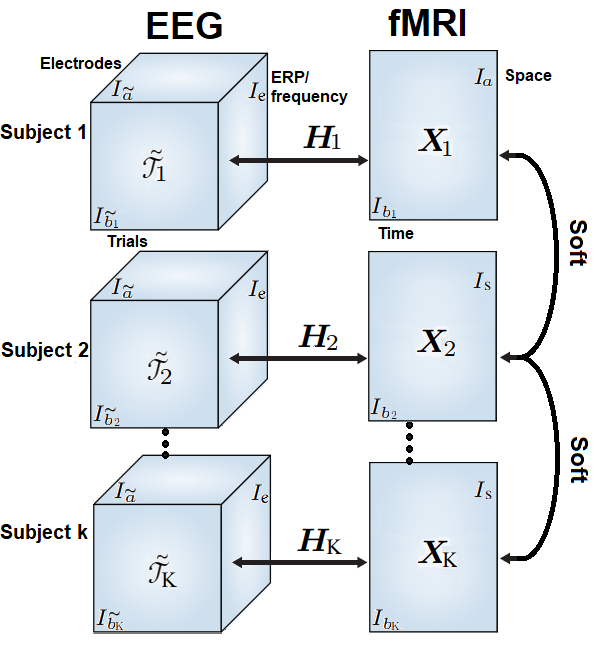}
    \caption{Schematic representation of DCMTF for $K$ subjects.}
    \label{dcmtf}
\end{figure}

Hence, in event-related designs as well as in studies like seizure detection, the different subjects can not be stacked in the same tensor since the multi-linearity assumption will certainly not be valid. Furthermore, a PARAFAC2 approach cannot be followed either, since the extra constraint of the constant cross product of PARAFAC2 is not valid. Although no connection among the time courses of the different subjects exist, similar areas are probably activated by similar stimuli (hence similar or same spatial maps). Hence, it is still beneficial to retain the neighborhood information exploited by the tensor formulation. In order to retain this multi-way structure (but still respect the difference in time courses) the formulation of the problem can be transformed to a Double (in time among EEG and fMRI and in space among subjects in fMRI) CMTF (DCMTF), as shown in Fig.~\ref{dcmtf}.

The 3rd-order EEG tensors, $\boldsymbol{\mathcal{\tilde{T}}}_k  \in \mathbb{R} ^{I_e \times I_{\tilde{a}}\times I_{\tilde{b}}}$, describe the variation over the spatial $(\boldsymbol{\tilde{a}}_{k_r})$, the temporal $(\boldsymbol{\tilde{b}}_{k_r})$ and the spectral/ERP $(\boldsymbol{e}_{k_r})$ modes, for $K$ different subjects. The fMRI matrices, $\boldsymbol{X}_k  \in \mathbb{R} ^{I_{a}\times I_{b}}$, contain the variation over the temporal $(\boldsymbol{b}_{k_r})$ and spatial $(\boldsymbol{a}_{k_r})$ modes, with the matrix $\boldsymbol{A}_k=\begin{bmatrix}\boldsymbol{a}_{k_1},\boldsymbol{a}_{k_2},\ldots,\boldsymbol{a}_{k_R}\end{bmatrix}$ comprising the weights of the $R$ spatial components of the $k$th subject and $\boldsymbol{A}$ being a spatial map with which all the subject spatial maps are similar (imposed through a regularization term). The parameter sets $\{z_k\}$ describe the subject-specific HRF matrix, $\boldsymbol{H}_k$, which will be optimized using either the double Gamma model~\cite{2017_Eyndhoven_HRF} or the Lennard-Jones model or any other appropriate model selected. The proposed cost function is given by:

\begin{align}
\label{hatzi3}
\sum_{k=1}^{K} ( \| \boldsymbol{\mathcal{\tilde{T}}}_k - \sum_{r=1}^{R} \boldsymbol{\tilde{a}}_{k_r} \circ  \boldsymbol{\tilde{b}}_{k_r}  \circ \boldsymbol{e}_{k_r} \|_F^2 +
\| \boldsymbol{X}_k &- \sum_{r=1}^{R} \boldsymbol{a}_{k_r} \circ ( \boldsymbol{H}_k (t,\{z_k\}) \boldsymbol{b}_{k_r}) \|_F^2 \nonumber \\ 
+ \lambda_1 \| \boldsymbol{A}_{k} - \boldsymbol{A} \|_F^2 ) 
\end{align}

For the coupling in the time domain, instead of using the flexible approximation with the subject-specific HRF, another soft coupling can be used and, hence, the cost function will become

\begin{align}
    \label{hatzi4}
\sum_{k=1}^{K} ( \| \boldsymbol{\mathcal{\tilde{T}}}_k - & \sum_{r=1}^{R} \boldsymbol{\tilde{a}}_{k_r} \circ  \boldsymbol{\tilde{b}}_{k_r}  \circ \boldsymbol{e}_{k_r} \|_F^2 +
\| \boldsymbol{X}_k- \sum_{r=1}^{R} (\boldsymbol{a}_{k_r} \circ  \boldsymbol{b}_{k_r})\|_F^2  \\  \nonumber
& + \lambda_1 \| \boldsymbol{A}_{k} - \boldsymbol{A} \|_F^2 + \lambda_2 \| \boldsymbol{B}_{k} - \boldsymbol{H\tilde{B}}_{k} \|_F^2) 
\end{align}

It should be noted that the tuning of two different $\lambda$ parameters might be difficult, but the decomposition of each subject separately, can provide information about the similarity of the spatial maps. A high value of $\lambda_1$ means that the spatial maps of all subject are the same and hence, hence imposing the same constraint in the spatial domain as Equation (12) (the assumption of the same spatial maps is implicitly made by the tensor decomposition introduced in Section~3). Despite the fact that matrices, not higher-order tensors, are considered, the coupling among the spatial components retains the multi-way nature of the multi-subject fMRI case (keep in mind that a 3-way tensor can also be represented as a set of matrices hard coupled in both of their modes~\cite{2013_sorber_optimization}), so the multi-way nature of the data is still exploited. The tuning of $\lambda_2$ is equivalent to $\lambda_B$ of Equation (12).

\section{Simulation results}

A simulated dataset similar to the one used in~\cite{2010_lei_steff} and~\cite{2014_dong_steff2} has been employed in our analysis. A disc with 2452 voxels (dipoles) was created in order to generate the data. For EEG, a concentric three-sphere model with 128 electrodes was set to wrap the disc, and the lead-field matrix computed in~\cite{2014_dong_steff2} has been used. The temporal sampling rate of EEG was 1 kHz while the epoch of the ERPs was set to 400 ms. The fMRI spatial maps were simulated as 2D images of $70 \times 70$ voxels, with the aid of the  SimTB~\cite{2012_erhardt_simtb} toolbox. In comparison to~\cite{2010_lei_steff,2014_dong_steff2}, the overlap in time for EEG and in space for fMRI has been increased.  In Fig.~5, the assumed neurophysiological sources can be viewed, from left to right: ``vision area'' S1, ``default mode network'' S2, ``auditory cortex'' S3, ``sensory networks'' S4, ``cognition areas'' S5 and ``dorsal attention network'' S6. The activity level at each active voxel was randomly sampled from a Uniform~[0.8,1.2] distribution for each replication of each simulation condition. These assumed active neural sources (rows a, b) along with the assumed ERPs (row d) yield scalp distributions and single-trial images in EEG and spatial maps and time courses of fMRI. Single-trial images (row c) are generated by multiplying each ERP (row d) with the trial amplitude (row a). Scalp potential distribution maps (topoplots, row e) are computed by solving the forward problem for each spatial map of row a. The fMRI BOLD signals (time courses, row f) were computed through the convolution between the trial amplitude (row a) with the canonical HRF. 
\emph{In all of the scenarios, we assume coupling of fMRI and EEG in the time domain only, hence $\lambda_A=\lambda_C=0$. Similar conclusions can be reached if the coupling is assumed in one of the other modes.}

This section will be split in 3 subsections:

\begin{itemize}
    \item We will exhibit the difference of the soft coupling approximation with the flexible approximation proposed in~\cite{2017_Eyndhoven_HRF} through a comparison study (based on Pearson correlation). Furthermore, in this subsection, we will study the tuning of the $\lambda$ value in the soft coupling method.
    \item Different methods will be examined in the case where all the subjects have the same time course: Parallel ICA, uncoupled CPDs (separately decomposing each tensor), hard and soft coupling in the time domain with different $\lambda_B$ values. 
    \item The same methods will be tested also in the last subsection, but different time courses per subject will be considered, in order to point out the need of an alternative formulation of the problem in such a case. 
\end{itemize}

The implementations of the proposed soft coupled decomposition and the DCMTF were performed within the Structured Data Fusion (SDF) framework~\cite{2015_sorber_structured} of Tensorlab~\cite{2016_vervliet_tensorlab} and Non Linear Squares (NLS) was adopted as the optimization scheme. Parallel ICA was implemented (using Group ICA) as in~\cite{2010_lei_steff,2015_hunyadi_parallel}, based on InfoMax~\cite{1995_bell_infomax} for the ICA step.

\begin{figure} 
  \centering
  \includegraphics[width=\linewidth]{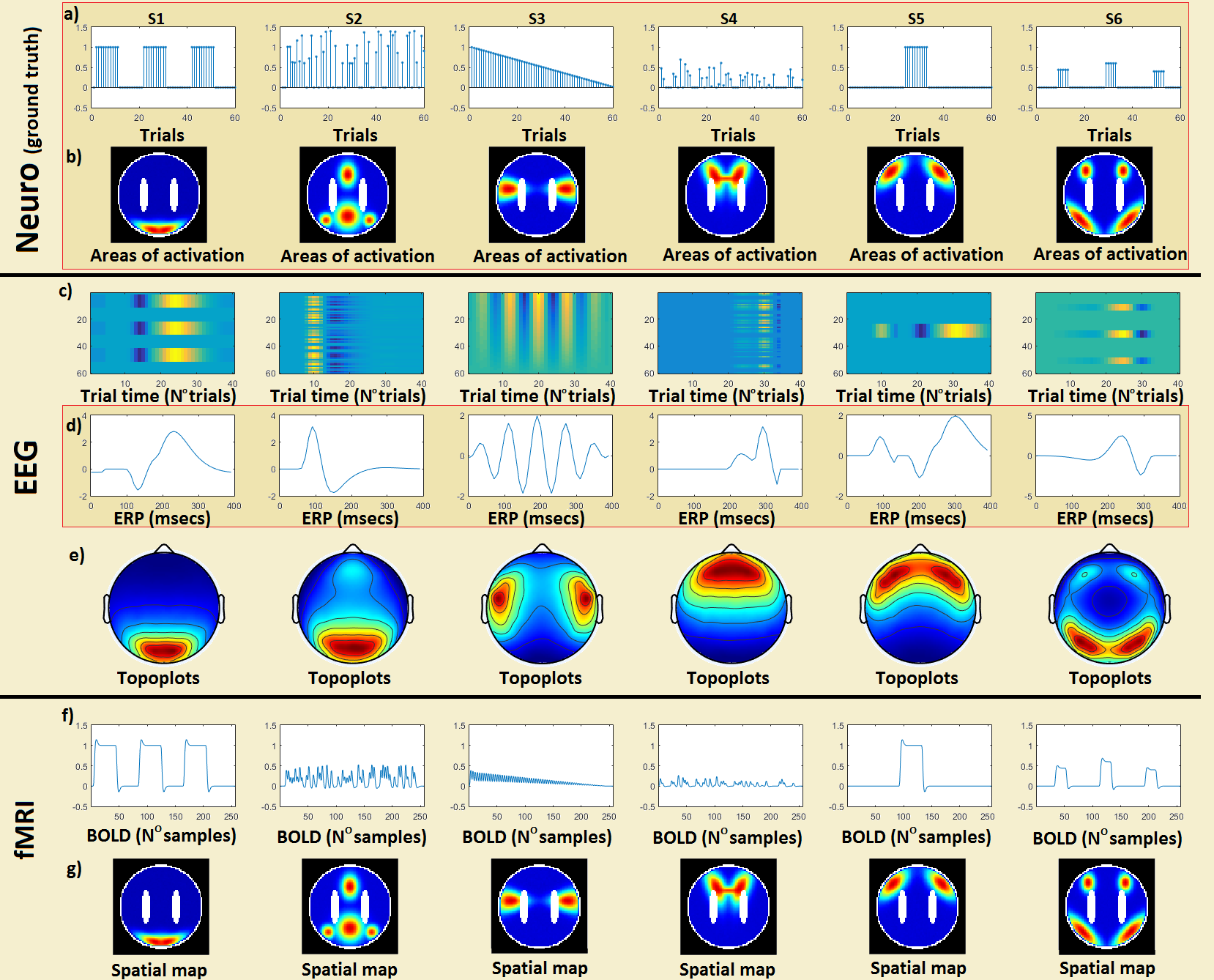}
  \caption{Simulated sources in EEG and fMRI.} 
  \label{fig:sim}
\end{figure}

In order to estimate $R_c$, $\boldsymbol{\mathcal{T}}$ and $\boldsymbol{\mathcal{\tilde{T}}}$ are separately decomposed, and a correlation matrix is computed based on the coupled modes of the tensors. Components with similarity exceeding a predefined threshold $t$ comprise the common components~\cite{2016_genicot_initfusion}. The computation of the number of the coupled components, $R_c$ could be incorporated in the cost function, similarly to~\cite{2017_acar_acmtf1,2017_acar_acmtf2}. 

In this way, we can also get an indication for the appropriate $\lambda$ values to be used: higher correlation indicates higher values for $\lambda$; hence the $\lambda$'s of the modes which will not be coupled will be set to zero. It is really important that the data of both modalities must be normalized (to unit norm) beforehand (so that the first two terms in (11) have the same weight in the cost function) and preprocessed for removal of artifacts~\cite{2018_walach_normalization}. 

The optimal initialization for each modality separately is not guaranteed to be the optimal one for their combination; furthermore, the permutation issues must be taken into consideration. Hence, an initialization method designed specifically for coupled decompositions must be used. For the initialization of the coupled tensor decomposition, the Generalized EigenValue Decomposition (GEVD)-based method proposed in~\cite{2015_sorensen_coupled} is used.\footnote{Special thanks to Nico Vervliet, KU Leuven, for sharing the code by M. S{\o}rensen, University of Virginia, USA.} When prior information is available for any of the modes (or part of them), the respective columns can be excluded from the optimization function and set equal (or almost equal) to the known factors.

Every experiment has been run 30 times (same map and time course, different activation amplitude and different instance of random noise each time). The Pearson correlation values presented in the following figures and tables are the mean Pearson correlation of all the obtained sources with the ground truth. Since the same algebraic initialization is used for every run, the standard deviations of all methods are relatively small, hence they will be reported only in the case that there are differences among the methods. 

\subsection{Soft versus flexible coupling}

We will compare the two alternative methods, that we will use to replace the hard coupling. Additionally, we will also examine the significance of the tuning of the $\lambda$ which controls how ``strong" the assumption of coupling will be.

\begin{figure}
  \centering
  \includegraphics[width=\linewidth]{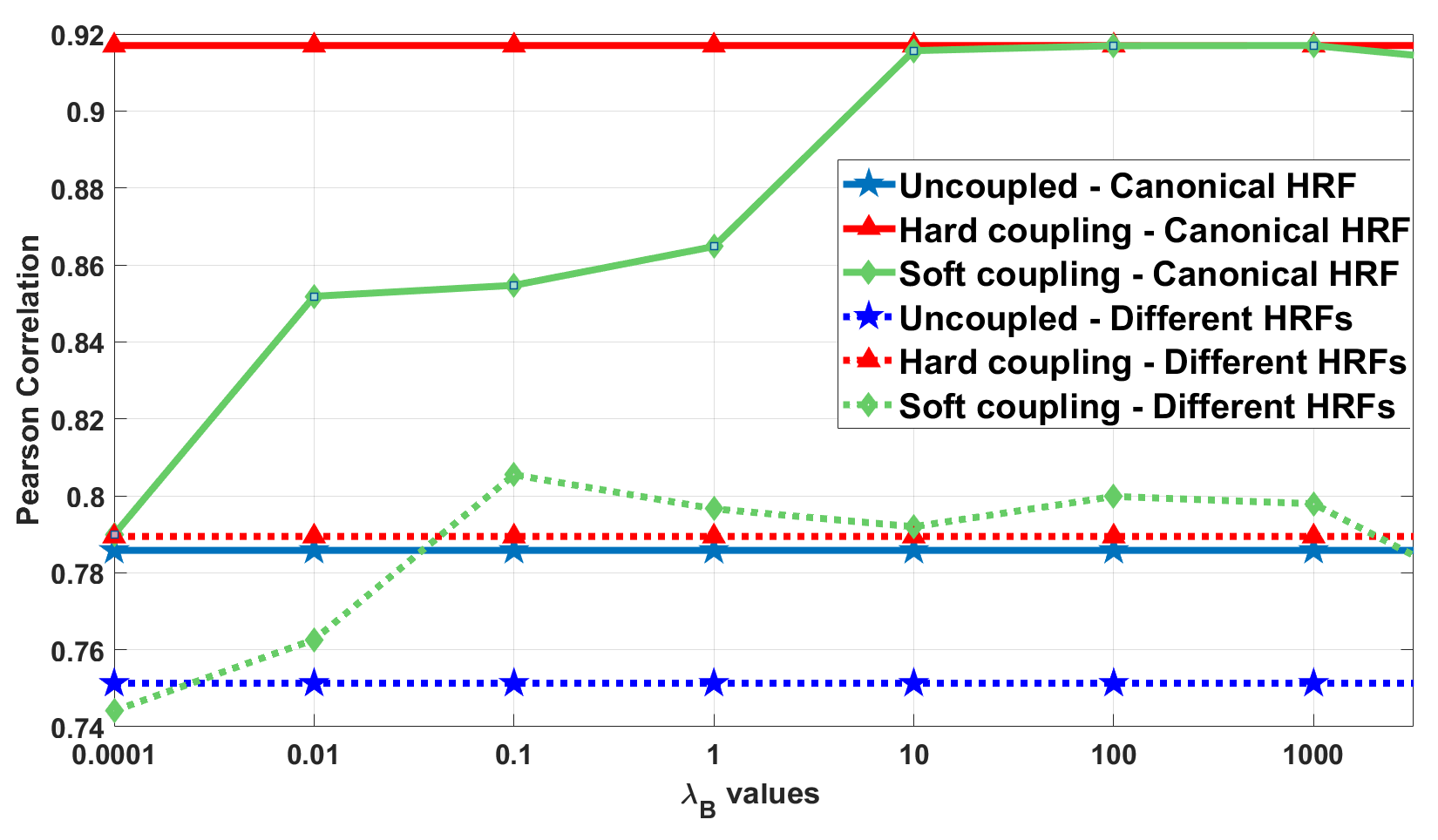}
  \caption{Correlation of the obtained sources with Uncoupled, Hard coupled and Soft coupled CPDs with different $\lambda_B$ values. } 
  \label{fig:lambda}
\end{figure}

Fig.~\ref{fig:lambda} visualizes the importance of the choice of the $\lambda_B$ value for soft coupling. We can distinguish two cases. In the first case (the solid lines), where the coupling assumption is exact (the simulated data were generated with the use of the canonical HRF, $\boldsymbol{H}$), it can be readily seen that the hard coupling is the best to use. However, the soft coupling analysis can reach the same performance with the appropriate tuning of $\lambda_B$. In the second case (dotted lines), the assumption of exact coupling is violated as the time courses were generated by convolution with different HRFs (the 5 different HRFs  presented in~\cite{2018_morante_info} have been used, which have a mean correlation of 0.8 with the canonical HRF), while $\boldsymbol{H}$ (Equation (6)) was constructed based on the canonical HRF. The fact that the time courses are similar but not equal deteriorates the performance of the hard coupling. Hard coupling still performs better than the uncoupled version but it is outperformed by the soft coupling for $\lambda_B > 0.1$. In cases where we move the HRF farther from the canonical HRF, the hard coupled case can become even worse than the uncoupled one.

\begin{figure}
  \centering
  \includegraphics[width=\linewidth]{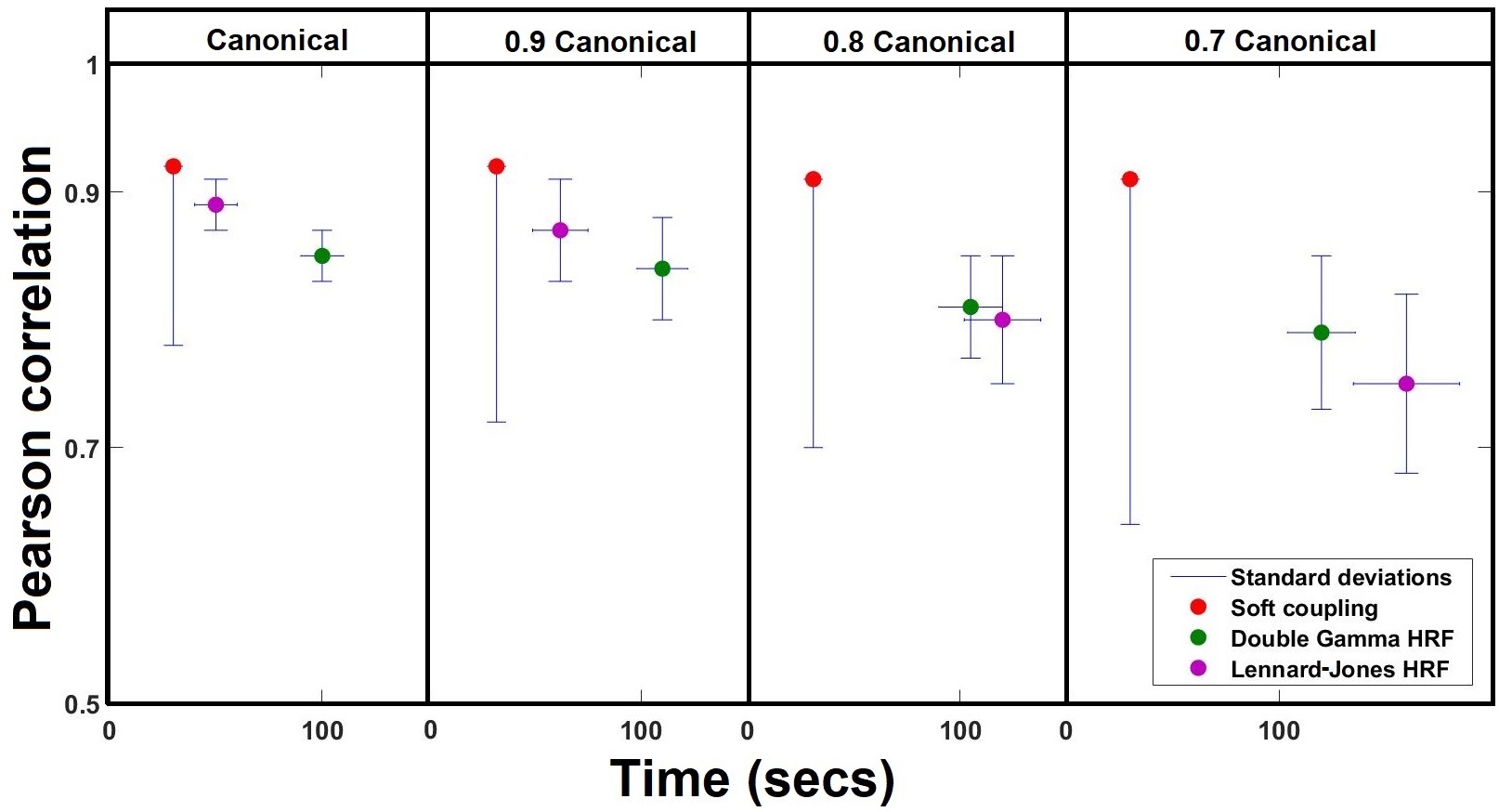}
  \caption{Comparison, based on Pearson correlation and time till convergence, of the soft coupling method with the flexible coupling method using the double Gamma HRF model and the Lennard-Jones HRF model. } 
  \label{fig:hrfs}
\end{figure}

For the comparison between the soft coupling and the flexible coupling, we simulated a similar single-subject scenario in order to test the performance and the computational burden for every method. We have simulated 4 different scenarios; in each scenario we slightly modify the HRF from which the data are generated. Initially we generated the data with the canonical HRF, while for the other scenarios an HRF with 0.9 correlation with a canonical one was used (0.8 and 0.7, respectively). We can see (Fig.~\ref{fig:hrfs}) that if we manage to tune appropriately the $\lambda_B$ value then the soft coupling method outperforms the other methods (red dot) but its deviation (suboptimal selection of $\lambda_B$ randomly chosen from $\{0.01,0.1,10,100,1000\}$) is large and its performance can be even worse than that of the flexible methods. It seems that the selection of the appropriate model (Lennard-Jones or double Gamma) is a compromise between accuracy and time complexity. In the cases where the HRF is closer to the canonical one, the Lennard-Jones model has similar performance as the double Hamma model in a significantly shorter time. In the cases where the HRF differs more from the canonical one, the performance deteriorates and the time needed to converge can also become longer (while also higher standard deviation is observed). It should be noted that the time needed for the selection of the $\lambda$ value is not represented in the figure since it depends on the intervals of the grid used in the grid search approach followed.

\subsection{Soft coupled tensor decompositions}

To compare the soft coupled tensor decompositions and the double coupled matrix tensor decomposition, multi-subject scenarios were simulated. The data from each subject contained all the six sources presented in Fig.~\ref{fig:sim} with different activation levels; the activation patterns have strengths randomly sampled from a Uniform~[2,5] distribution. Five different subjects are simulated, and for the simulations presented in this section each subject is assumed to have the same time course for every source (differing only in the noise) while in the simulation used in the next section differences are incorporated in the time courses and HRFs of some of the subjects. 

\begin{figure}
  \centering
  \includegraphics[width=.9\linewidth]{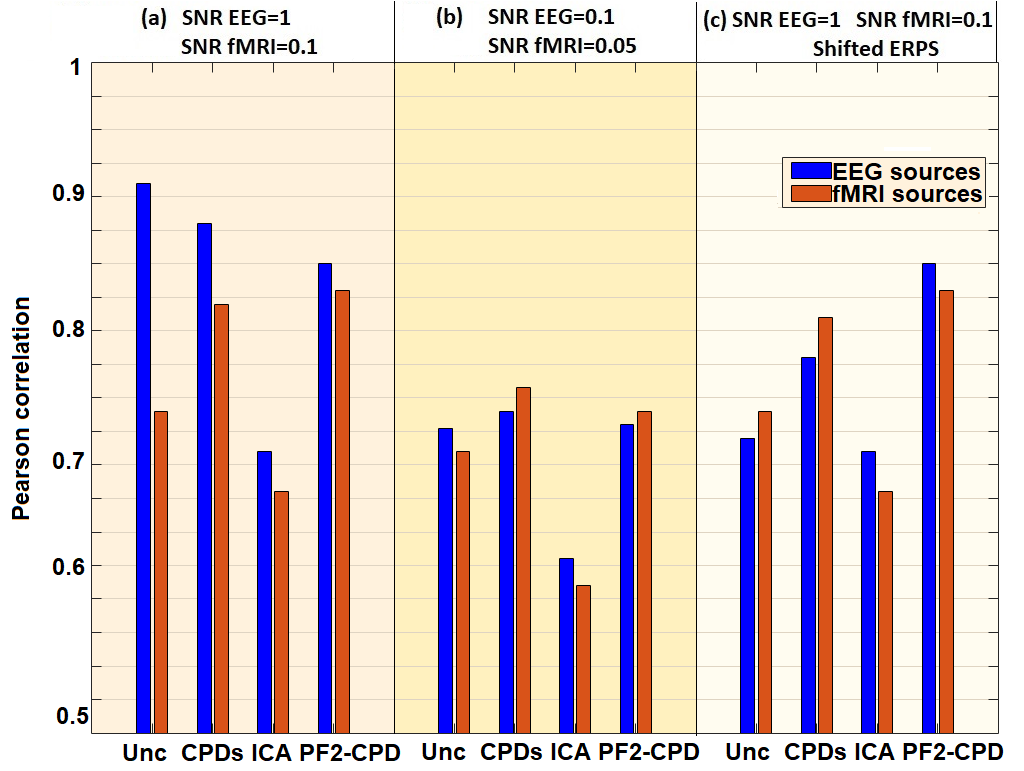}
  \caption{Accuracy of different methods. EEG: diamonds, fMRI: discs} 
  \label{fig:ticks}
\end{figure}

\begin{figure} [b]
  \centering
  \includegraphics[width=\linewidth]{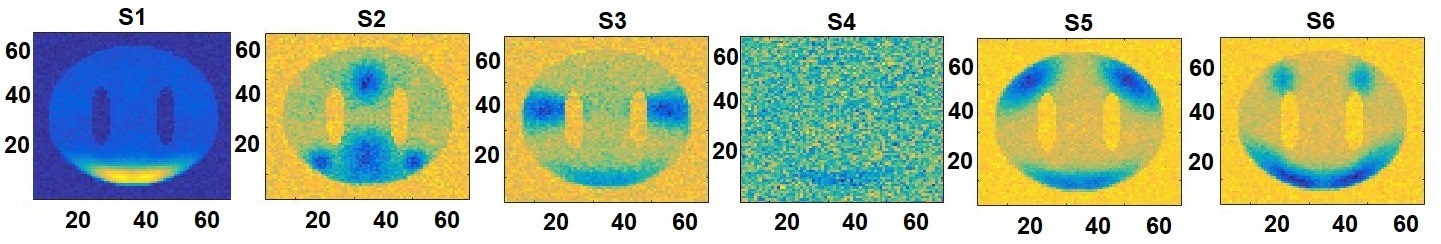}
  \caption{Resulting fMRI spatial maps with ICA, at SNR=0.1.} 
  \label{fig:ica}
\end{figure}

In Fig.~\ref{fig:ticks}, the mean correlation between the obtained sources and the ground truth per method and per modality (diamonds for EEG and discs for fMRI) at different Signal to Noise Ratios (SNR= squared Frobenius norm of the signal over the squared Frobenius norm of the noise) can be observed. In cases (a) (same noise level as in~\cite{2010_lei_steff}) and (b), different levels of noise are tested, while in case (c) the assumption of the same ERP per subject is violated and the ERPs are shifted (the first subject has 0 msec shift while subjects 2--5 have time shifts at increments of 10~msecs with respect to the 1st subject, hence a shift of 40~voxels in the 5th subject). Parallel ICA exhibits inferior performance compared to both the uncoupled (Unc) and soft coupling methods (Coupled CPDs, ``CPDs'' and Coupled PARAFAC2 CPD, ``PF2-CPD'') in all of the cases, due to the overlapping in the sources, which violates the independence assumption. The resulting spatial maps obtained by spatial ICA in case (a) can be viewed in Fig.~~\ref{fig:ica}. Note that, in the areas of overlapping, there is crosstalk between the maps. S4, which overlaps with most of the rest of the sources, can not be identified (for comparison with the ground truth, observe row g of Fig.~\ref{fig:sim}). It can be seen that, in case (a), the correlation for EEG with uncoupled analysis is higher than with soft coupling. This is caused by the performance gain for the fMRI source in the coupled case which results in a slight loss for EEG. Overall, the correlation is increased with soft coupling. In case (b), where the SNR is the same for both modalities, soft coupling yields better results. PF2-CPD in both (a) and (b) cases yields a slightly worse result than coupled CPDs (since the multilinearity assumption used by CPD is valid here). The last case, (c), is the one where the advantage of PARAFAC2 becomes apparent. We observe that ICA is affected by the ERP shifting much less than the uncoupled and the coupled CPD methods, but it still has the worst performance.

\subsection{Flexible double coupled matrix tensor decomposition}

In this subsection, we will test the case where subject variability exists, in time or in space. Hence three different scenarios and two subscenarios (for each) will be tested: In the first scenario, all the subjects have the same time courses (similar to previous subsection). In the second scenario, each subject has a different HRF and a time shift in each of the time courses of the sources. The 5 different HRFs presented in~\cite{2018_morante_info} have been used while also the time courses are shifted (the first subject has no shift while subjects 2--5 have time shifts at increments of 1~sec with respect to the 1st subject, hence a shift of 4~secs in the 5th subject). In the last scenario, the time courses of the subjects are the same but the spatial maps of every subject are different: subject variability was introduced in the spatial domain of two of the sources (2 and 4) and rotation (in increments of 4~degrees per subject) of one of the sources (2) and voxel shifts at increments of 2 voxels with respect to the 1st subject in the other source (4).

In every scenario, two subscenarios are also simulated: a) Only sources 2, 3 and 4 (Fig.~\ref{fig:sim}) with low spatial overlap are used, and b) all the sources are used. We considered these subscenarios in order to examine the impact of overlap, time shift and subject spatial variability, separately.

\begin{table}
 \centering
\begin{adjustbox}{width=.98\linewidth}
\begin{tabular}{ | c | c || c | c || c | c|| c | c| }
    \hline
\multicolumn{2}{|c||}{\multirow{2}{*}{\textbf{Methods}}} & \multicolumn{2}{|c||} {\textbf{Same time and space} } & \multicolumn{2}{|c||} {\textbf{Diff. time courses}} & \multicolumn{2}{|c|} {\textbf{Diff. spatial maps}} \\
\hhline{~~------}
\multicolumn{2}{|c||} {} & {Low overlap} & {High overlap}  & {Low overlap} & {High overlap} & {Low overlap} & {High overlap} \\
    \hline
      \hline
 \multicolumn{2}{|c||} {Parallel ICA}  & \textbf{0.95} $\pm$ 0.02 &  0.80 $\pm$ 0.02 & 0.82 $\pm$ 0.02 & 0.68 $\pm$ 0.11 & \textbf{0.88} $\pm$ 0.02 & 0.75 $\pm$ 0.02\\
 \multicolumn{2}{|c||} {Uncoupled} & 0.85 $\pm$ 0.02 &  0.82 $\pm$ 0.02 & 0.70 $\pm$ 0.6 & 0.69 $\pm$ 0.08 & 0.70 $\pm$ 0.09 & 0.69 $\pm$ 0.10\\
 \multicolumn{2}{|c||} {Coupled Tensors} & \textbf{0.95} $\pm$ 0.02& \textbf{0.92} $\pm$ 0.02& 0.75 $\pm$ 0.02& 0.65 $\pm$ 0.03& 0.78 $\pm$ 0.02 & 0.70 $\pm$ 0.03\\
 \multicolumn{2}{|c||} {DCMTF} &  0.91 $\pm$ 0.03& 0.90 $\pm$ 0.03 & \textbf{0.90}  $\pm$ 0.04 & \textbf{0.90} $\pm$ 0.03 & \textbf{0.88}  $\pm$ 0.04& \textbf{0.87} $\pm$ 0.04\\
  \hline  
   \end{tabular}
   \end{adjustbox}
    \caption{Performance of the different fusion methods under different scenarios.}
    \label{table:tab2}   
\end{table}

The mean Pearson correlation of the obtained sources with the ground truth is given in Table~\ref{table:tab2} for the two scenarios. It can be noted that the Parallel ICA method outperforms the other methods in the case where no severe spatial overlap and the same time courses per subject exist. However (as mentioned previously), this method is also affected more severely by the spatial overlap of the sources (since the assumption of the joint distribution of the sources is violated) and additionally it is affected by different time courses per subject since it is based on the assumptions imposed by Group ICA (GICA)~\cite{2005_beckmann_tensorial}. In the case of high overlap and same time course per subject, the soft coupled tensor decomposition exhibits the best performance but on the other hand this method is most affected by the differences in the time courses, since the assumption of multi-linearity is violated in the time domain. The Uncoupled tensors have similar behaviour since the difference in the time course per subject remains even if the EEG and fMRI tensors are decomposed separately. The DCMTF model allows the  successful estimation of the underlying sources much better than the other methods in the case of different time courses and HRFs per subject. It should be noted that the performance of DCMTF is similar to that of Soft Coupled Tensors in the case of the same HRF and time course per subject.  With different spatial maps per subject, we can note that the ICA-based method is affected less since the assumption of same time course used by GICA is then valid. Concerning the standard deviation of the Pearson correlation we can note that GICA is the more stable method with slightly higher standard deviation in the cases where it fails. 

The tuning of $\lambda_1$ is less significant than that of $\lambda_2$. The performance of DCMTF presented in Table~1 in the first two scenarios is with a high value of $\lambda_1$ ($\lambda_1=10^6$), since the spatial maps per subject are the same, while for the last scenario the $\lambda_1$ value was selected based on grid search.

\subsection{Discussion}

From the results obtained in the previous subsections we can understand that if the correct model is selected (based on the type of the problem at hand), the use of ``non-hard'' (soft or flexible) coupling methods and raw data can improve the obtained result. 

In every case, the method has to be selected a-priori by the user based on the type of problem. An initial analysis of the data of both modalities separately is recommended. This initial analysis can provide relevant information to the user regarding the model to be selected as well as hints on the values of the (hyper)parameters.

For example, a simple mean-ERP analysis prior to the joint analysis could provide an indication of the amount of the shift in the ERPs, in order to select which of the soft coupled tensor decompositions should be used (PARAFAC2-CPD or coupled CPDs). As mentioned previously, also the number of coupled components, $R_c$, can be obtained by setting a threshold based on the correlation among the components of the two modalities in the initial ``separate'' decomposition. 

Furthermore, it should be kept in mind that the design of the experiment plays a significant role in the model selection. An experiment with different time course per subject will lead to the adoption of  DCMTF; this does not mean that the multi-way nature of the data will still not be exploited (as previously noted), since the coupling among the spatial modes of fMRI will enrich the optimization problem with spatial neighborhood information among the different subjects.

Concerning reproducibility of the results, we have noted that all the methods have a low standard deviation provided they succeed to correctly separate the sources (high mean Pearson correlation). We have also noted that although the standard deviation of the correlation is low in the Uncoupled tensors method in cases where it it fails, the standard deviation of the correlation among the estimated sources in every run is high (can reach up to 0.40). This means that although the method produces 
(almost) equally bad results in every run (since the assumption is not valid) the results differ from one run to another (when the method fails). On the other hand, the coupled tensors method, though it also fails when different time courses exist, produces similarly bad results in every run. This difference could be possibly explained from the coupling constraints which enhance the uniqueness properties of the decomposition.

We have demonstrated that the use of raw data in the problem of fusion of EEG and fMRI, provided the heterogeneity of the data variables~\cite{2018_walach_normalization}  is carefully handled, facilitates accurate source identification. As it has been pointed out~\cite{2015_Lahat_Multimodal,2015_Adali_Fusionb}, the use of the raw data can improve the result of the decomposition by exploiting latent correlations between the different datasets, which might have been attenuated by the use of intermediate feature extraction methods (such as GLM). Our findings (especially those of Section~5.3) confirm the inability of GLM (and hence all methods relying on GLM as a preprocessing step, e.g., Parallel ICA) to cope with HRF variability~\cite{2009_lindquist_hrf,2014_swinnen_jica}. Moreover, we have confirmed that ICA-based methods fail to correctly decompose overlapped sources~\cite{2007_stegeman_comparing,2019_chatzichristos_journal}.

\section{Conclusions}

This pre-print briefly reviews the literature of the problem of EEG-fMRI fusion and reports our recent results on this topic, which are based on  the adoption of two different tensor models for jointly analyzing fMRI and EEG data. This is an attempt to benefit from the multi-way nature of \emph{both} modalities, performing an early fusion, and hence, bypassing the need to rely on features.  Performance gains have been reported compared to ICA methods as well as to the separate analyses of the datasets. The use of coupled PARAFAC2-CPD was seen to outperform the coupled CPD in the presence of shifts in the ERPs per subject. A comparison between flexible and soft coupling approaches has been presented while also an alternative HRF model has been tested for the first time.  Future work will include  studies with real data, comparisons with methods based on Independent Vector Analysis (IVA)~\cite{2015_Adali_Fusiona} and alternative tensor models (e.g., Block Term Decomposition~\cite{2017_chatzichristos_BTD}). Moreover, a more systematic selection of the $\lambda$ values will be
sought for.

\section*{Acknowledgment}
The authors would like to thank Dr. Li Dong, UEST, China for providing the lead-field matrix used in~\cite{2014_dong_steff2}, Dr. Loukianos Spyrou, Univ. of Edinburgh, UK, Dr. Nico Vervliet and Simon Van Eyndhoven, KU Leuven, Belgium for fruitful discussions on the topics of EEG,  soft and flexible coupling in Tensorlab, respectively and Manuel Morante Moreno, University of Athens, for the cooperation in the Lennard-Jones model. Furthermore, we would like to thank Prof. M. Davies and J. Escudero, Univ. of Edinburgh, UK  who were coauthors in the conference paper~\cite{2018_chatzichristos_fusion}, a preliminary version of this work. The research leading to these results was funded by the European Union's $7^{\mathrm{th}}$ Framework Program under the ERC Advanced Grant: BIOTENSORS ($n^{\circ}~339804$). This work was also funded by EU H2020 MSCA-ITN-2018: On integrating Magnetic Resonance SPectroscopy and Multimodal Imaging for Research and Education in MEDicine (INSPiRE-MED) Grant Agreement $n^{\circ}~339804$. This research also received funding from the Flemish Government (AI Research Program).

\bibliographystyle{IEEEbib}
\bibliography{Template} 

\end{document}